\documentclass[twocolumn,superscriptaddress,amsmath,amssymb,showpacs,showkeys]{revtex4}
\usepackage{graphicx,color}% Include figure files
\usepackage{dcolumn}% Align table columns on decimal point
\usepackage{bm}% bold math 
\newcommand{\beq}{\begin{equation}} \newcommand{\eeq}{\end{equation}}
\newcommand{\bqa}{\begin{eqnarray}} \newcommand{\eqa}{\end{eqnarray}}

\definecolor{gold}{rgb}{0.75,0.56,0.00}
\definecolor{green}{rgb}{0.00,0.50,0.00}

\begin{document}

\preprint{APS/123-QED}

\title{Quantum filtering one bit at a time}

\author{Jason F. Ralph, Neil P. Oxtoby}
 \email{jfralph@liv.ac.uk, n.oxtoby@liv.ac.uk} 
  \affiliation{ Department of Electrical Engineering and Electronics, The University of Liverpool,\\  Brownlow Hill, Liverpool, L69 3GJ, United Kingdom.}
 
\date{\today}

\begin{abstract} 
In this paper we consider the purification of a quantum state using the information obtained from a continuous measurement record, where the classical measurement record is digitized to a single bit per measurement after the measurements have been made. Analysis indicates that efficient and reliable state purification is achievable for one- and two-qubit systems. We also consider quantum feedback control based on the discrete one-bit measurement sequences. 
\end{abstract}

\pacs{03.65.Wj, 03.65.Yz}

\keywords{quantum state-estimation, continuous weak measurement}

\maketitle

%------------------------------------------------------------------------------------------------------------------------------------------------------%
%   INTRODUCTION  %
%------------------------------------------------------------------------------------------------------------------------------------------------------%

In classical systems, filtering a sequence of measurements in order to reconstruct the behavior of a classical object is well established~\cite{Track01}. The measurements represent information that has been obtained about the current state and dynamical evolution of the system. The state constructed using this data represents the `best guess' of the state of the system {\em given} the measurement record. This state-estimation process is often referred to as {\em tracking}~\cite{Track01}. In quantum systems, an equivalent approach has been developed using continuous weak measurements~\cite{Bel1980,WisemanMilburnBook}.  This approach has shown a range of interesting results using the reconstructed (or partially reconstructed) quantum state to apply controls to modify the subsequent quantum evolution \cite{FeedbackFiltering01}. Of particular interest here is rapid state purification introduced by Jacobs \cite{Jac03} and later studied by other authors \cite{RSP1,RSP2,RSP3}. In rapid state purification, the rate at which a state can be purified can be modified by applying simple controls. 

A major difference between classical state-estimation filters and quantum filtering is that, in quantum systems, the act of measurement has an irreducible effect on the system being measured. A projective (von Neumann-type) measurement is an extreme example, leaving the quantum system in an eigenstate of the measured operator. In continuous weak measurement models, the quantum system of interest is considered to be coupled to an environment. Projective measurements are taken on the environment, which contains information on the system of interest through the coupling. By averaging over the environmental degrees of freedom, it is possible to form an equation for the evolution of the quantum system of interest, which contains Hamiltonian evolution that has been perturbed by the effect of the environmental measurements. The average evolution of the quantum system in the presence of the environment is given by the master equation, and the particular evolution for a {\em given} set of measurements (a realization) is represented by a stochastic master equation (SME)~\cite{WisemanMilburnBook}. The conditioned evolution for a given measurement record is often referred to as an `unraveling' of the master equation~\cite{WisemanMilburnBook}. Such unravelings may be continuous or discontinuous (`quantum jumps'), depending on the nature of the interaction and the measurement process. They represent quantum filtering processes similar to a classical tracking filter. The measurement record generated by a continuous weak measurement is a classical analog signal. The signal is typically very noisy (quantum noise generated by the back action from the measurement interaction) and integration over discrete time steps allows the SME to be calculated numerically to reconstruct the quantum state.

This paper is motivated by two factors. Firstly, it has been shown recently that for a discontinuous/jump unravelling of the master equation, there is a minimum number of classical states that are required to track the state of a quantum system with $D$-energy levels \cite{Kar11}. In the case of a qubit, it was demonstrated that a single classical bit (i.e. two classical states) was sufficient to track the qubit's state -- that is, if one knew the classical state at any point in time, one would be able to infer the state of the qubit as it jumps between states. Secondly, it is known in signal processing that if a classical signal is sent through a number of parallel (thresholded) channels, each of which is corrupted by Gaussian noise, then the optimum number of thresholds that can be used to discretize this signal reduces to one as the noise level increases (i.e. one bit per channel) \cite{McD06}. This is related to a classical nonlinear effect: supra-threshold stochastic resonance \cite{Sto00}. 

In the cases considered here, parallel channels (i.e. multiple measurement interactions) would introduce additional noise through the measurement back action. Instead, we use the fact that the underlying signal is slowly varying with respect to the measurement/purification rate to perform an (implicit) temporal average over the noise. The time required to complete the purification is dependent on the coupling to the environment (the measurement strength), which is far longer than the incremental measurements (or samples) used in the SME. We show that a single threshold (a one-bit record, OBR) in the SME is sufficient to purify the state of one or two qubits, with or without feedback control based on local unitary operations. Crucially, the thresholding of the signal is performed {\em after} the classical measurement record is generated so it does not add to the back action on the quantum system. We demonstrate that the fidelity of one-bit filtering can be very good, reproducing classical and quantum correlations between qubits (as measured by the quantum discord \cite{discord}). The importance of these results rests in the potential for experimental realization of quantum filtering -- reducing a continuous time analog signal to a discrete-time single bit measurement sequence provides a significant reduction in the data required to reconstruct the quantum state of the system of interest.

%------------------------------------------------------------------------------------------------------------------------------------------------------%
%      QUANTUM FEEDBACK AND CONTROL
%------------------------------------------------------------------------------------------------------------------------------------------------------%

In Jacobs' rapid state purification~\cite{Jac03}, a qubit starts in a completely mixed state (i.e. an unknown state) and undergoes controlled rotations to rotate its Bloch vector onto the plane orthogonal to the measurement axis. These controlled rotations alter the rate of increase of the average purity by a factor that approaches two as the purity of the state $P\rightarrow 1$ deterministically (where $P(\rho) = Tr(\rho^2)$). Alternatively, if feedback is used in the opposite manner, to rotate the Bloch vector towards the measurement axis, purification is stochastic and the time required to reach a given purity can be reduced relative to Jacobs' protocol~\cite{RSP1}. (Jacobs considered discrete `bang-bang' controls~\cite{Jac03}, but this is distinct from the discretized measurement and filtering process discussed here).

In this paper, we use two stochastic master equations: one SME to generate a measurement record and another to estimate the state of the system conditioned on the measurement record. The first SME (SME1) is initialized in a pure state and remains in a pure state ($\rho_{0}$). If an experiment were available, SME1 would be redundant, its main role is to generate the measurement record. The secondary role of SME1 is to provide a reference against which we can compare our estimate of the state. SME2 generates the estimated state. SME2 is initialized in a completely mixed state, which will purify gradually as information is extracted. All feedback controls are based on SME2 and the estimated state only. 

Continuous weak measurements are not restricted to Hermitian operators but -- for simplicity -- we will assume that the measurement operator is Hermitian. For Hermitian weak measurements $\hat{c}_{j}  = \sqrt{(2k_{j}\hbar)}\hat{y}_{j}  = \hat{c}_{j} ^{\dagger}$, where $k_{j} $ is the strength of the $j$th measurement interaction, the stochastic master equation (SME1) for the density matrix, $\rho_{0}$, is given by~\cite{WisemanMilburnBook},
\begin{eqnarray}\label{sme1}
d\rho_0&=&-\frac{i}{\hbar}\left[\hat{H},\rho_0\right]dt
-\sum_{j} k_{j}\left[\hat{y}_{j},\left[\hat{y}_{j},\rho_0 \right]  \right]dt \nonumber\\
&&+ \sum_{j} \sqrt{2k_{j}}\left(\hat{y}_{j}\rho_0+\rho_0\hat{y}_{j}-2\left\langle\hat{y}_{j}\right\rangle_0 \rho_0 \right)dW_{j}\end{eqnarray}
where $\hat{H}$ is the Hamiltonian of the system, $dt$ is an infinitesimal time increment and  $dW_{j} $ is a real Wiener increment (such that $dW_{j}=0$ and $dW_{j} dW_{j'}  = \delta_{jj'}dt$). The corresponding continuous measurement records are 
\begin{equation}\label{meas1}
\frac{dy_{j}(t)}{\sqrt{\hbar}} = \sqrt{8k_{j} }\left\langle \hat{y}_{j}  \right\rangle_0 dt + dW_{j}
\end{equation}
To solve this problem numerically, finite time steps $\Delta t$ are integrated (Euler method) to find a discrete measurement $\Delta y_{j}(t)$. This is a classical analog signal, which represents the sampled signal that would be derived from an experiment. For the one-bit measurement record, the discrete samples are replaced with,
$$
\Delta Y_{j}(t) = \sqrt{(\hbar \Delta t)} sgn(\Delta y_j)
$$
where $sgn(x)=\pm 1$ depending on the sign of $x$. The estimated state, $\rho_{c}$,  is constructed using SME2,
\begin{eqnarray}\label{sme2}
\Delta\rho_{c}&=&-\frac{i}{\hbar}\left[\hat{H},\rho_{c}\right] \Delta t
-\sum_{j} k_{j}\left[\hat{y}_{j},\left[\hat{y}_{j},\rho_{c} \right]  \right] \Delta t \nonumber\\
&&+ \sum_{j} \sqrt{2k_{j}}\left(\hat{y}_{j}\rho_{c}+\rho_{c}\hat{y}_{j}-2\left\langle\hat{y}_{j}\right\rangle_{c} \rho_{c} \right) \Delta W_{j}
\end{eqnarray}
where $\Delta W_{j} = \Delta Y_{j}(t)/\sqrt{\hbar}- \sqrt{8k_{j} }\left\langle \hat{y}_{j}  \right\rangle_c \Delta t  $. This is analogous to the classical {\em innovation}, the difference between the actual measurement and that predicted using the estimated state~\cite{Track01}.

Here we will consider two specific cases: one qubit with a single measurement along the $Z$-axis, and two coupled qubits with a measurement of $ZZ = \sigma_z \otimes \sigma_z$ (where $\sigma_r$ is a Pauli matrix with $r = x,y,z$). In each case, we will assume that the qubits have a background Hamiltonian, which corresponds to a rotation about the $X$-axis with an oscillation frequency $\omega_0$. Such Hamiltonians are common in solid state qubits, where they correspond to tunneling interactions. Local controls are used because they are seen as a tractable way to preserve entanglement for quantum processing applications~\cite{localcontrols}. We will use the fidelity $F= F(\rho_{0},\rho_{c}) = \left|{\rm Tr}\left[\sqrt{\sqrt{\rho_{c}} \rho_{0} \sqrt{\rho_{c}}}\right]\right|^2$~\cite{Peters04} to characterize the errors in our estimate $\rho_{c}$ of the true state $\rho_{0}$. For the two qubit case, we calculate classical correlations ${\cal C}(\rho)$~\cite{discord},
$$
{\cal C}(\rho) = S(\rho^B)-\begin{array}{c} \min \\ \{\Pi^A_a\} \end{array} \sum_a p_a S(\rho^B|A_{\{\Pi^A_a\}})
$$
and the quantum discord ${\cal D}(\rho) = S(\rho^A)+S(\rho^B)-S(\rho)-{\cal C}(\rho)$~\cite{discord}, where $S(\rho)$ is the von Neumann entropy $S(\rho) = -Tr(\rho \log_2 \rho)$ and $\rho^A$ and $\rho^B$ are the reduced density matrices for the individual qubits, and the minimum is taken over the set of projectors on qubit $A$, $\{\Pi^A_a\}$. $S(\rho^B|A_{\{\Pi^A_a\}})$ is the entropy of qubit $B$ under the action of the projectors on qubit $A$, and $p_a$ is the probability associated with the result obtained for that projector on qubit $A$. Orthogonal projectors are used here because, although the minimum value that they give is not always the true minimum, the value of the quantum discord is very close to the true value~\cite{discord2}.   
\begin{figure}[htbp]\label{Fig_1}
	\centering
	        \includegraphics[width=0.5\textwidth]{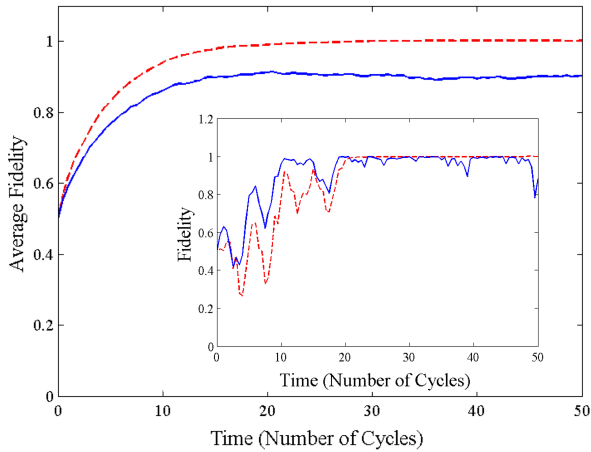}
		%\vspace{3cm}
	\caption{\label{fig:Average Fidelity} (Color online) Average fidelity of one qubit as a function of time with no controls: full record (red-dash) and OBR (blue-solid). Inset: an example run, showing fidelity for one measurement record.}
\end{figure}

%------------------------------------------------------------------------------------------------------------------------------------------------------%
%      ONE QUBIT PURIFICATION
%------------------------------------------------------------------------------------------------------------------------------------------------------%
\begin{figure}[htbp]\label{Fig_2}
	\centering
		\includegraphics[width=0.5\textwidth]{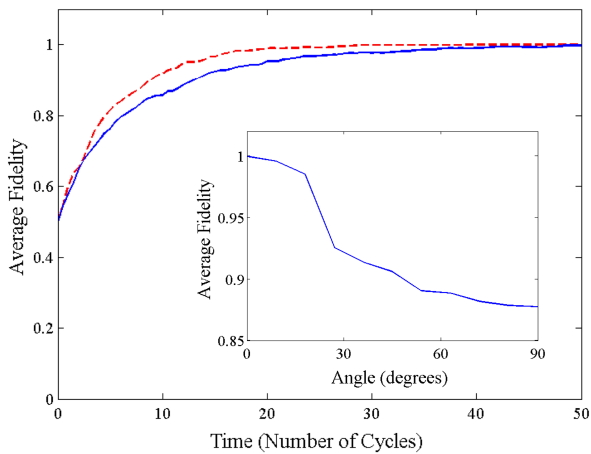}
		%\vspace{3cm}
	\caption{\label{fig:Average Fidelity} (Color online) Average fidelity of one qubit as a function of time with controls applied to rotate Bloch vector onto measurement axis: full record (red-dash) and OBR (blue-solid). Inset: Average fidelity after 50 cycles as a function of the angle to the measurement axis (in $X$-$Z$ plane) for OBR.}
\end{figure}

For the single qubit case we have a Hamiltonian $H = \hbar \omega_0\sigma_x/2$, an initial state $\rho_c(0)=I/2$ and a single measurement operator $\hat{c}_1 = \sqrt{(2k_1\hbar)}\sigma_z$. In Figure 1, we show the average fidelity as a function of time for the case where no controls are applied and the single-qubit Bloch vector rotates freely about the $X$-axis. The (angular) frequency $\omega_0$ is fixed and other parameters are scaled relative to $\omega_0$: $k_1=0.005\omega_0$. SME2 is integrated numerically using the Euler method with $10000$ steps per cycle and averaged over 500 measurement records. In the limit $\Delta t\rightarrow dt$, the first two moments of  the OBR and the full record are identical (mean zero, variance $\hbar dt$) and it is possible to show that the two records will purify at the same rate, on average. The analytical results in~\cite{RSP1} also apply for the OBR: the average time to approach unit purity is one-quarter of the time to reach an average purity of unity. For finite time steps the OBR purifies slightly slower than the full record (approx. 20-30\% longer). However, it is clear that, although the SME using the OBR does estimate the underlying state of the system reasonably well, the average fidelity for the state generated by the OBR saturates at a fixed level $< 1$. In each run of the SME, the fidelity approaches one but suffers from occasional drops in fidelity as the errors in the state estimated by the OBR accumulate, see Figure 1(inset). Approximately five to ten thousand OBR sample points are required per qubit cycle, longer time steps can make the filter unstable, with purities growing significantly above one. The number of sample points required for a stable filter is related to the qubit frequency and the measurement strength. For larger $k$ values, more frequent sample points are needed but purification takes less time in total.

To improve the fidelity of the final state estimate it is necessary to apply feedback controls to rotate the Bloch vector towards the measurement axis at each time step. This is the feedback protocol studied in \cite{RSP1}. Figure 2 shows the effect of adding the feedback control. The fidelity of the final state for the OBR approaches one as the system purifies, albeit slightly slower than for the full record. This is true whether there is Hamiltonian evolution or not. In the case where there is no Hamiltonian evolution, the protocol in \cite{RSP1} reduces to a classical (no feedback) measurement, but in either case, the OBR allows the state to be purified and the fidelity of the final state approaches one in the long time limit. The inset in Figure 2 shows the dependence of the final (average) fidelity on the angle between the measurement axis and the vector towards which the Bloch vector is rotated. This shows that the fidelity of the final purified state is relatively insensitive to small errors in the feedback controls -- the final fidelity is very close to one for all angles $< 20^o$.

%------------------------------------------------------------------------------------------------------------------------------------------------------%
%      TWO QUBIT PURIFICATION
%------------------------------------------------------------------------------------------------------------------------------------------------------%

\begin{figure}[htbp]\label{Fig_3}
	\centering
		\includegraphics[width=0.45\textwidth]{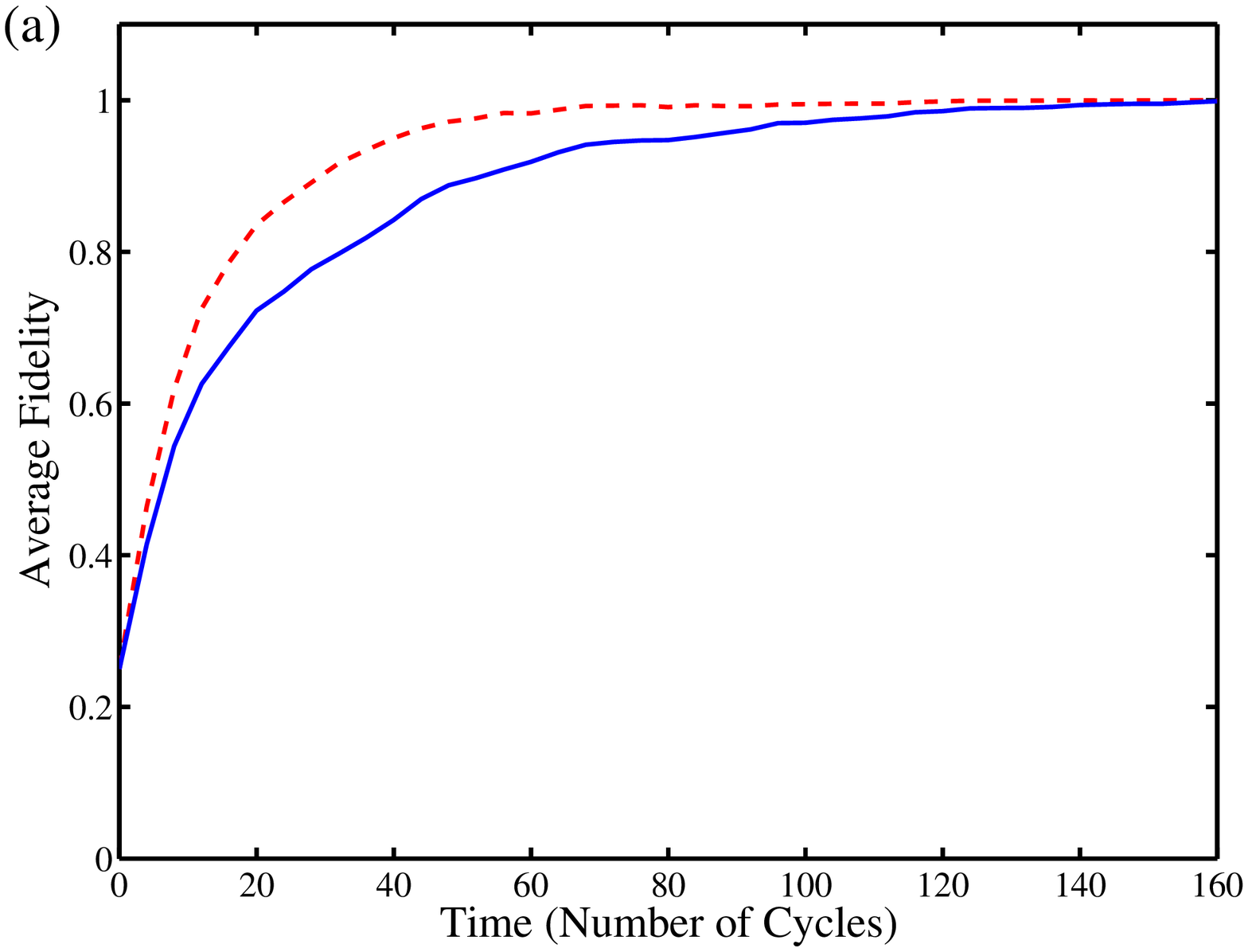}
		\includegraphics[width=0.45\textwidth]{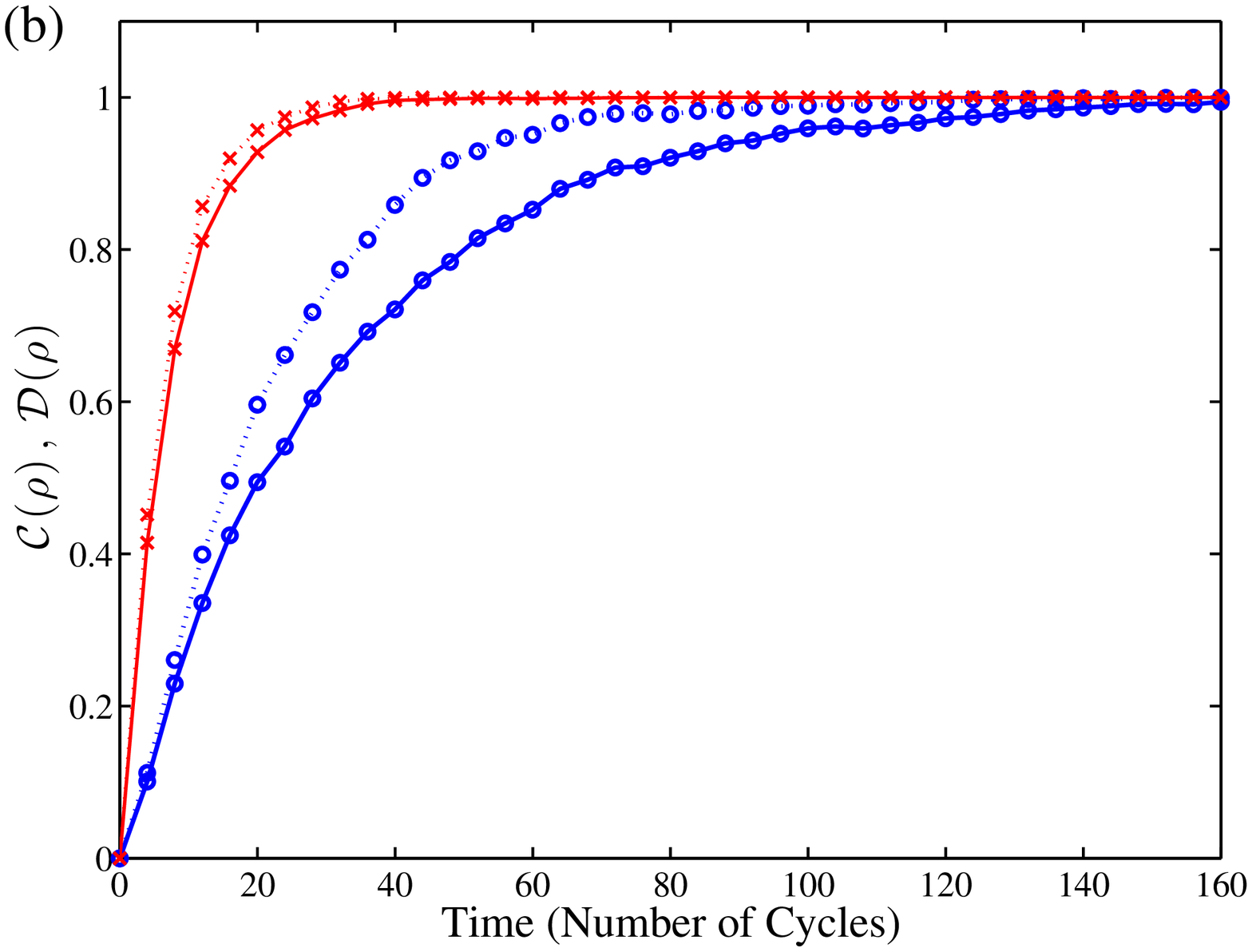}
		%\vspace{6cm}
	\caption{\label{fig:Average Fidelity} (Color online) (a) Average fidelity of the two qubit state as a function of time with single $ZZ$ measurement record and no controls: full record (red-dash) and OBR (blue-solid), and (b) classical correlations (red-cross) and quantum discord (blue-circle) of the two qubit state as a function of time with single $ZZ$ measurement record and no controls: full record (dot) and OBR (solid).}
\end{figure}

For two qubits, the situation is more complex with the need to estimate the combined state and (ideally) to retain the ability to generate entangled states. Starting with a two qubit Hamiltonian, $H_{A+B} = \hbar \omega_0(\sigma_x\otimes I)/2+\hbar \omega_0(I \otimes\sigma_x)/2$, there are two main ways to generate an entangled state from a separable initial state: an entangling Hamiltonian or an entangling measurement. An interaction of the form $H_{int} =\kappa(\sigma_z\otimes\sigma_z)$ can generate entanglement. When combined with two local $Z$ measurements and no controls, the full record purifies the state effectively but the OBR generates an unstable state estimate which occasionally drifts away from the true state, as with the example shown in Figure 1. Using local controls for the individual (reduced) density operators, rotating onto the respective $Z$-axes using either record will purify the two-qubit state with a fidelity that tends to one in the long time limit. However, this has the disadvantage of removing entanglement between the two qubits, because the resultant state is separable with each qubit in an eigenstate of $Z$. The alternative approach is considered here, where entanglement is generated by the measurement. This is similar to the case in \cite{RSP3}, where a $ZZ$ measurement was used to purify and entangle two qubits. In \cite{RSP3}, the two qubits were separated into two encoded qubits and an analogue of Jacobs' protocol based on local operations was used to purify and to protect entanglement by manipulating the system into a decoherence free subspace. Here, Jacobs' protocol would purify the combined state at the expense of the fidelity of the final state -- as seen in the case of the single qubit above -- and the alternative approach based on \cite{RSP1} is not available with purely local operations. Instead, we opt to use no feedback in this case, relying on the inherent $X$-rotations and a single (entangling) $ZZ$-measurement. Taking an initial state, $\rho_c(0)=(I\otimes I)/4$, $k_1=0.005\omega_0$, and using 20000 steps per cycle (the same data rate per qubit as in the previous example -- and subject to the same requirement on data rate for a stable filter), the purity of the estimated state approaches one for both the full record and the OBR. The OBR purifies the state slower than the full record, but the fidelity does tend to one in the long time limit. The OBR also captures the classical and quantum correlations very well (see Figure 3(b)). Both the full record and the OBR reproduce the classical correlations very rapidly, but the quantum correlations indicated by the quantum discord, ${\cal D}(\rho)$, are more difficult to estimate using the OBR.

%------------------------------------------------------------------------------------------------------------------------------------------------------%
%        CONCLUSIONS
%------------------------------------------------------------------------------------------------------------------------------------------------------%

In this paper, we have demonstrated that quantum filtering can be performed using a one bit measurement record. Single qubit states and two qubit entangled states undergoing continuous weak measurement can be estimated accurately from a single bit per time step. This represents a significant reduction in the data required to use quantum filtering methods with only minor reductions in filter performance. The one-bit measurement record can provide an accurate state estimate for one qubit when local feedback is applied, and it has been shown that the estimates are insensitive to small feedback errors. For two qubits, it is possible to purify the combined state with local measurements and local feedback, but this is at the expense of entangled states. If an entangling measurement is used, it is possible to use the one-bit  record without feedback to generate entangled states with fidelities approaching one. The one-bit record also provides accurate estimates of the classical and quantum correlations.

\begin{center}
	\textit{Acknowledgments}
\end{center}
\noindent  The authors would like to thank Dr Mil\'{e} Gu (National University of Singapore) for helpful and informative discussions in the preparation of this paper. The authors acknowledge the support of an ESPRC grant: EP/G007918. 

\bibliographystyle{apsrev}

\end{document}